# Non-Mie Optical Resonances in Anisotropic Biomineral Nanoparticles

Roman E. Noskov[1,2,*], Ivan I. Shishkin[1,2], Hani Barhom[1,2] and Pavel Ginzburg[1,2]

[1] *Department of Electrical Engineering, Tel Aviv University, Ramat Aviv, Tel Aviv 69978, Israel*
[2] *Light-Matter Interaction Centre, Tel Aviv University, Tel Aviv, 69978, Israel*

**Abstract:** Optical properties of nanoparticles attract continuous attention owing to their high fundamental and applied importance across many disciplines. Recently emerged field of all-dielectric nanophotonics employs optically induced electric and magnetic Mie resonances in dielectric nanoparticles with a high refractive index. This property allows obtaining additional valuable degrees of freedom to control optical responses of nanophotonic structures. Here we propose a conceptually distinct approach towards reaching optical resonances in dielectric nanoparticles. We show that lacking conventional Mie resonances, low-index nanoparticles can exhibit a novel anisotropy-induced family of non-Mie eigenmodes. Specifically, we investigate light interactions with calcite and vaterite nanospheres and compare them with the Mie scattering by a fused silica sphere. Having close permittivities and same dimensions, these particles exhibit significantly different scattering behavior owing to their internal structure. While a fused silica sphere does not demonstrate any spectral features, the uniaxial structure of the permittivity tensor for calcite and non-diagonal permittivity tensor for vaterite result in a set of distinguishable peaks in scattering spectra. Multipole decomposition and eigenmode analysis reveal that these peaks are associated with a new family of electric and magnetic resonances. We identify magnetic dipole modes of ordinary, extraordinary and hybrid polarization as well as complex electric dipole resonances, featuring a significant toroidal electric dipole moment. As both vaterite and calcite are biominerals, naturally synthesized and exploited by a variety of living organisms, our results provide an indispensable toolbox for understanding and elucidation of mechanisms behind optical functionalities of true biological systems.

[*]Corresponding author, e-mail: nanometa@gmail.com

Subwavelength light-matter interactions have been long associated with plasmonic structures made of metals such as gold or silver[1]. However, significant ohmic losses and questionable biocompatibility have restricted areas of their immediate and potential applications, nevertheless some groundbreaking approaches have been developed[2,3]. Metallic spheres can support a family of electric-type resonances (dipole, quadrupole, etc.) due to light coupling with the surface plasmonic oscillations. At the same time, their vanishing inductance leads to negligible magnetic responses. In order to reach a considerable inductance along with magnetic resonances, additional geometrical degrees of freedom should be involved, as it was demonstrated with split-ring resonators[4], particles clusters (e.g. dimers[5]), necklaces[6], or raspberry-like metamolecules[7], constitutive parts of metamaterials[8,9], and many others.

As a promising alternative to plasmonic structures, a new branch of nanophotonics has emerged and aimed at employing optically induced Mie resonances in nanoparticles with a high refractive index[10]. Dielectric particles provide an alternative route toward high-Q optical resonances. According to Mie theory, the scattering properties of spherical particles consisting of isotropic and nonmagnetic materials depend only on two parameters: the particle radius $R$ and the light wavelength inside the particle $\lambda/n$ ($n$ is the particle refractive index, the background index is 1)[11]. Hence, both electric- and magnetic-type resonances (Mie resonances) occur for the wavelengths at which light exhibits constructive interference after an integer number of wave roundtrips inside the particle, i.e., the ratio $2R/(\lambda/n)$ remains fixed for a given eigenmode[10]. In order to meet this condition, the particle should be either relatively large in size (i.e. comparable with the *free space* wavelength) or have a high refractive index. The latter also allows strong energy confinement into the subwavelength volume. In practice, Mie resonances become well-distinguished for $n>2$, their scattering efficiency is gradually increasing with the growth of the refractive index, and the linewidth of spectral peaks is reducing[12,13]. This has given rise to demonstration of many promising optical effects, including magnetic hot spots generation[14,15], magnetic mirrors[16], highly directional Huygens-like nanoantennas[17–19], toroidal moments[20] (also demonstrated in plasmonics[21]), control over magnetic dipolar emission[22,23], bianisotropy[24,25], high numerical aperture lenses[26], etc[27].

Commonly employed materials with a high refractive index include semiconductors, such as titanium dioxide[28], germanium[29], silicon[30,31], gallium arsenide[32], and others[10]. Being favorable for nanoelectronics applications, nanoparticles consisting of these materials are typically characterized by high toxicity and low biocompatibility[33]. This does not allow extension of low-loss and heating-free Mie resonances and related phenomena into a set of biomedical applications.

The striking examples of materials playing an important role in biology, bio-organic chemistry and nanomedicine are polymorphs of calcium carbonate ($CaCO_3$): calcite and vaterite. Being the most stable polymorph of $CaCO_3$, calcite represents an optically transparent uniaxial crystal. Calcite microcrystals enable living organisms to achieve specific functions, ranging from skeleton support[34] to guiding and focusing light inside tissue[35,36]. Otoliths, cornerstones of the vestibular system of vertebrates, are made of calcite[37]. Vaterite is a metastable phase of $CaCO_3$, which eventually undergoes phase transformation into calcite under standard environmental conditions. Vaterite crystals can form polycrystalline spherical microparticles, also referred to as spherulites. Their strong porosity gives rise to highly efficient bioactive substance incorporation and targeted drug delivery[38–41].

To date, the studies of optical properties for calcite and vaterite micro- and nanoparticles have been predominantly concentrated on optomechanical manipulations[42,43]. Although, a number of approaches for calculation of electromagnetic scattering from anisotropic particles have been proposed[44–46], systematic analysis of optical resonances supported by the particles has been performed only for a rather simple case of radially symmetric anisotropy[47,48].

Here we comprehensively analyze optical properties of calcite and vaterite nanoparticles and reveal their complexity. Relatively low ordinary and extraordinary refractive indices (~1.4-1.7) make these structures unable to support conventional Mie eigenmodes. However, strong birefringence leads to a novel anisotropy-induced family of *non-Mie* optical resonances. By using the finite element method (FEM), we investigate light scattering spectra of calcite and vaterite spheres and demonstrate their tremendous differences from the scattering by an isotropic fused silica sphere with a close value of the refractive index. In the case of identically sized particles, both vaterite and calcite show a variety of distinct peaks, while fused silica is lacking any spectral features. Field decomposition with respect to spherical multipoles reveals that these new peaks are associated with electric- and magnetic-type resonances. Magnetic dipole modes of ordinary, extraordinary and hybrid polarization as well as complex electric dipole resonances in which the streamlines for the electric polarization can form closed loops as well as helicoidal-like structures are demonstrated for the first time. The analysis of Cartesian electric and magnetic moments show that the revealed non-Mie electric resonances feature a significant toroidal electric dipole moment.

**RESULTS AND DISCUSSION**

Figure 1 summarizes main material properties of fused silica, calcite and vaterite. Fused silica is an isotropic optical glass with the optical permittivity $\varepsilon_s \approx 2.13$. Calcite is a negative uniaxial crystal ($\varepsilon_{eo} < \varepsilon_o$), characterized by ordinary and extraordinary permittivities $\varepsilon_{xx} = \varepsilon_{yy} = \varepsilon_o \approx 2.75$ and $\varepsilon_{zz} = \varepsilon_{eo} \approx 2.2$[49]. Here we consider vaterite in the form of a spherical spherulite. Such structure contains single nanocrystal subunits that arrange as a bundle of fibers tied together at the center and spread out at the ends (so-called a "dumbbell" or "heap of wheat" model)[50,51]. Each subunit is a positive uniaxial monocrystal ($\varepsilon_{eo} > \varepsilon_o$) with $\varepsilon_o \approx 2.4$ and $\varepsilon_{eo} \approx 2.72$[52,53]. The distribution of unity optical axes of such particles could be well approximated by the family of co-focused hyperboles, symmetrically rotated with respect to *z*-axis [51,54], as shown in Fig.1(c). The focal position $d$ is assumed to be controlled by the synthesis procedure. In the further analysis we set $d = R/2$, while this value can be adjusted by considering a specific fabrication protocol.

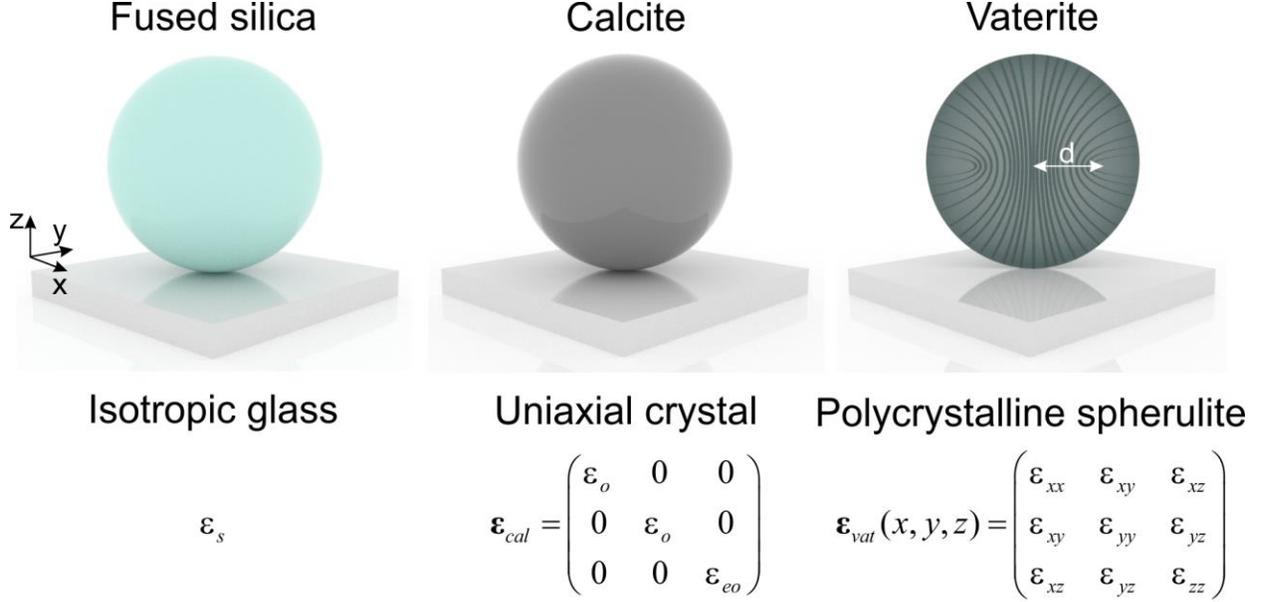

**Fig. 1:** Artistic view of fused silica, calcite, and vaterite spheres along with their permittivities. Fused silica is isotropic glass. Calcite is a positive uniaxial crystal. Vaterite spherulite represents densely packed co-focused hyperbolic fibers of single nanocrystal subunits, and *d* is the distance between hyperbolas focus and the center of the particle. Vaterite optical properties can be characterized in terms of the effective symmetric non-diagonal permittivity tensor whose all components are position-dependent (see the text for details).

Hyperbolic distribution of optical axes, high density of fibers and a small size of vaterite nanocrystal-based subunits (~ 25-35 nm[55]) in comparison to the light wavelength allows considering spherulite properties in the framework of the effective permittivity tensor, which can be calculated by expressing the local permittivity of a uniaxial subunit $\boldsymbol{\varepsilon}_{loc}$ into the global coordinate system, related with the geometrical center of a spherulite, i.e. $\boldsymbol{\varepsilon}_{vat} = A_z(\varphi)A_y(\psi)\boldsymbol{\varepsilon}_{loc}A_y^{-1}(\psi)A_z^{-1}(\varphi)$, where

$$\boldsymbol{\varepsilon}_{loc} = \begin{pmatrix} \varepsilon_o & 0 & 0 \\ 0 & \varepsilon_o & 0 \\ 0 & 0 & \varepsilon_{eo} \end{pmatrix},$$

while $A_y(\psi)$ and $A_z(\varphi)$ are rotation matrixes around *y*- and *z*-axis, respectively, $\varphi$ is the spherical azimuthal angle and $\psi$ is the angle between *z*-axis and the tangent line at the local point of a hyperbola. By accomplishing this procedure, the symmetric non-diagonal permittivity tensor can be derived[51,54]

$$\boldsymbol{\varepsilon}_{vat}(x, y, z) = \begin{pmatrix} \varepsilon_{xx} & \varepsilon_{xy} & \varepsilon_{xz} \\ \varepsilon_{xy} & \varepsilon_{yy} & \varepsilon_{yz} \\ \varepsilon_{xz} & \varepsilon_{yz} & \varepsilon_{zz} \end{pmatrix}, \quad (1)$$

where

$$\begin{aligned}
\varepsilon_{xx} &= \varepsilon_o\left[\cos^2\varphi\left(\cos^2\psi + \eta\sin^2\psi\right) + \sin^2\varphi\right], \\
\varepsilon_{yy} &= \varepsilon_o\left[\sin^2\varphi\left(\cos^2\psi + \eta\sin^2\psi\right) + \cos^2\varphi\right], \\
\varepsilon_{zz} &= \varepsilon_o\left[\sin^2\psi + \eta\cos^2\psi\right], \\
\varepsilon_{xy} &= \varepsilon_{yx} = \varepsilon_o\sin\varphi\cos\varphi\left[\cos^2\psi + \eta\sin^2\psi - 1\right], \\
\varepsilon_{xz} &= \varepsilon_{zx} = \varepsilon_o(\eta-1)\cos\varphi\sin\psi\cos\psi, \\
\varepsilon_{yz} &= \varepsilon_{zy} = \varepsilon_o(\eta-1)\sin\varphi\sin\psi\cos\psi,
\end{aligned}$$

and $\eta = \varepsilon_{eo}/\varepsilon_o$. The equations of a hyperbola and its tangent make this model to be self-consistent.

$$\begin{cases} \cot\psi = \dfrac{b^2}{a^2}\tan\vartheta \\ r = \dfrac{ab}{\sqrt{(a^2+b^2)\sin^2\vartheta - a^2}} \end{cases}, \qquad (2)$$

where $a$ and $b$ are semi-major and semi-minor axes ($d^2 = a^2 + b^2$), $r$ and $\vartheta$ are the spherical coordinates. By variations in $a$ and $b$ (which physically mean filling the whole spherulite cross-section with hyperbolic fibers), one can get an implicit parametric function $\psi(\vartheta, r)$, rendering all components of the spherulite permittivity tensor coordinate-dependent, as shown in Fig. 2.

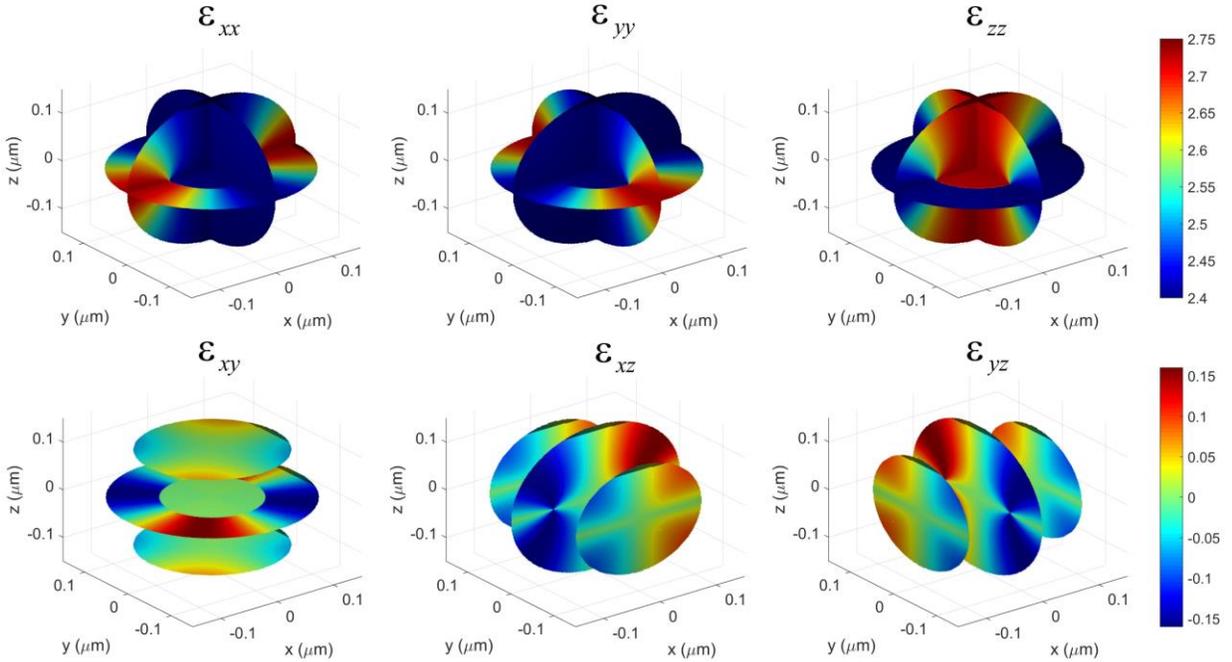

**Fig. 2:** Spatial structure of the effective permittivity tensor components for a vaterite spherulite with $R=150$ nm and $d=75$ nm, given by Eqs. (1) and (2).

The diagonal terms of the tensor obtain values between $\varepsilon_o$ and $\varepsilon_{eo}$, attributed to the vaterite single nanocrystal components; while non-diagonal terms vary between $\pm(\varepsilon_{eo}-\varepsilon_o)/2$. It is worth noting, that negative values of non-diagonal terms appear as a result of permittivity tensors of vaterite subunits representation in the global Cartesian coordinate system, which does not coincide with the local orientation of their optical axes. The inner region of the particle (can be considered as a quasi-cylindrical core) is defined by the hyperbolas focus positions ($x^2+y^2 \leq d^2$) and features pronounced Cartesian anisotropy, whereas the outer shell is radially anisotropic (i.e., characterized by different radial and azimuthal/polar components of the permittivity tensor in the spherical coordinate system), as the vaterite fibers radially diverge from the particle center. Thus, from an intuitive viewpoint on material properties, vaterite spherulites combine radial and Cartesian anisotropy (Fig. 2).

Next, we employ FEM analysis, implemented within CST Microwave Studio, in order to investigate scattering of a plane optical wave, impinging on fused silica, calcite and vaterite spheres. To provide quantitative characterization for the electromagnetic properties of the particles, we employ multipole expansions in terms of spherical multipoles[56]. This approach allows us to get the direct comparison with the classical Mie solution and underline pronounced differences.

We start with presenting the total scattering cross-section as a sum of contributions from different spherical multipoles, which are valid for an arbitrary form of the polarization current $\mathbf{J}(r)$:

$$C_{sca} = \frac{\pi}{k^2}\sum_{l=1}^{\infty}\sum_{m=-l}^{l}(2l+1)\left(|a_e(l,m)|^2 + |a_m(l,m)|^2\right), \qquad (3)$$

where the scattering coefficients are given by

$$a_e(l,m) = \frac{(-i)^{l-1}k}{E_0\sqrt{\varepsilon_h \pi(2l+1)(l+1)}}\iiint Y_{l,m}^*(\vartheta,\varphi)j_l(kr)\left\{k^2\mathbf{r}\cdot\mathbf{J}(r)+\left(2+r\frac{d}{dr}\right)\nabla\cdot\mathbf{J}(r)\right\}dV,$$

$$a_m(l,m) = \frac{(-i)^{l-1}k^2}{E_0\sqrt{\varepsilon_h \pi(2l+1)(l+1)}}\iiint Y_{l,m}^*(\vartheta,\varphi)j_l(kr)\left\{\mathbf{r}\cdot\nabla\times\mathbf{J}(r)\right\}dV.$$

Here the integration is performed over the particle volume, $E_0$ is the incident electric field amplitude, $\mathbf{r}$ is the radius vector, $\varepsilon_h$ is the permittivity of the host media, $k=(\omega/c)\sqrt{\varepsilon_h}$, $c$ is the speed of light in a vacuum, $Y_{l,m}(\vartheta,\varphi)$ is the scalar spherical harmonic, $j_l(kr)$ is the Bessel function of $l$th order. The polarization current can be expressed through the local electric polarization as $\mathbf{J}=\partial\mathbf{P}/\partial t = -i\omega\mathbf{P}$. The harmonic time dependence $\exp(-i\omega t)$ is assumed. For an isotropic spherical particle the scattering coefficients can be expressed analytically - those are well-known Mie solutions[11]. However, in a general case of an arbitrary particle shape, inhomogeneity and anisotropy, one should use Eq. (3) to extract the contributions of individual spherical multipoles to the total scattering cross-section for a given distribution of $\mathbf{P}$, which is calculated numerically.

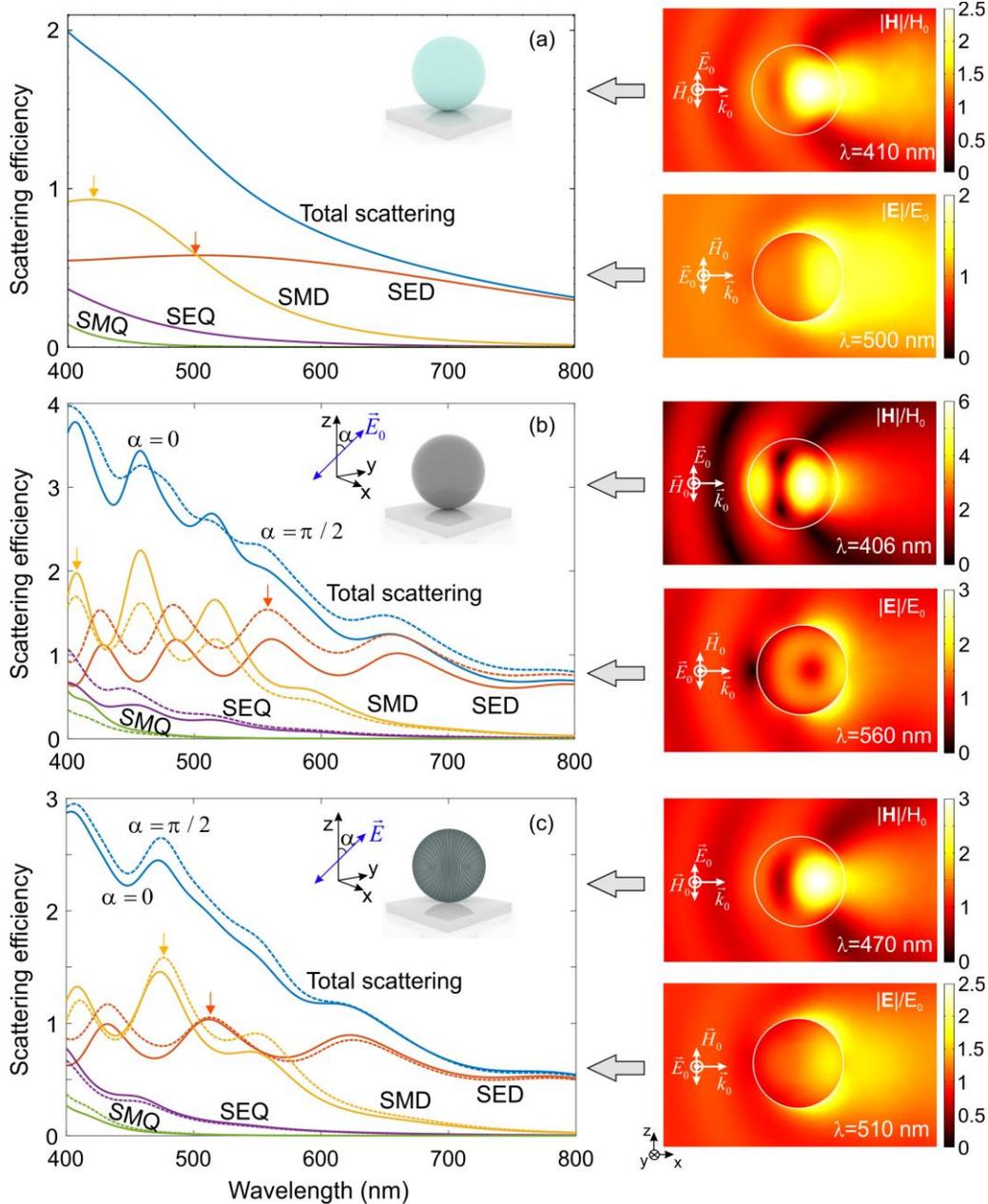

**Fig. 3:** Scattering cross-section normalized by the geometrical particle cross-section (scattering efficiency) vs wavelength for (a) fused silica, (b) calcite, and (c) vaterite. Contributions from spherical electric dipole (SED), magnetic dipole (SMD), electric quadrupole (SEQ), and magnetic quadrupole (SMQ) are distinguished with the spherical multipole decomposition (see the text). Continuous and dashed curves correspond to parallel and perpendicular orientation of the electric field with respect to $z$-axis. Yellow and red arrows indicate selected wavelengths for which the electric and magnetic field distributions (normalized by the electric $E_0$ and magnetic $H_0$ amplitude of the incident plane wave) are calculated, shown in the right panels (the coordinate system refers to all plots). Particles' radii are $R=150$ nm. Substrates in the insets are present for the illustration purposes only and are not a part of the numerical analysis.

We assume that the plane wave is propagating along $x$-axis, as shown in Fig. 3. Anisotropic properties of both calcite and vaterite particles allow us to define the axial symmetry with respect to *z*-axis. Therefore, to fully characterize scattering properties, one can decompose the electric field polarization to parallel and perpendicular components with respect to *z*-axis. In our consideration we take the comparatively small particle radii *R*=150 nm to avoid appearance of conventional Mie resonances in the optical domain. Remarkably, this range of sizes is of a particular practical interest, as it allows vaterite and calcite particles to penetrate through cell membranes, providing numerous opportunities for nanomedical applications[39,57].

Figure 3 shows the scattering efficiency spectra (the total scattering cross-section normalized by the geometrical particle cross-section) and individual contributions from different spherical multipoles in the visible spectral domain. The fused silica particle does not show any pronounced Mie resonances because of a relatively low permittivity contrast with the background. In contrast, the spectra of calcite and vaterite spheres are characterized by well-distinguished resonant peaks, associated with excitation of electric and magnetic dipole modes. This is also evidenced by observing electric and magnetic field structures, shown in Fig. 3 for the selected wavelengths corresponding to the resonances of spherical electric and magnetic dipoles. Being very weak for a glass particle, the field confinement inside the particle is considerably increasing in cases of calcite and vaterite. Hence, one may conclude that these peaks are caused by a novel type of optical resonances induced by particle's anisotropy. This behavior has a new nature, which is fundamentally different from high refractive index-related retardation effects or surface charge (plasmonic) oscillations.

To elucidate these anisotropy-driven resonances in detail, we employ numerical eigenmode solver, which reveals the field structure of electric- and magnetic-type modes. The results of eigenmode decomposition for calcite and vaterite nanoparticles are shown in Figs. 4 and 5, respectively. The distributions for both electric and magnetic fields underline the core of the effect, provided by their anisotropy. Specifically, the magnetic modes at $\lambda = 478$ nm and $\lambda = 556$ nm for calcite (Figs 4 (b) and (c)) and $\lambda = 480$ nm and $\lambda = 555$ nm for vaterite (Figs. 5 (a) and (b)) demonstrate the structure, which is similar to the magnetic dipole given by Mie solution – namely the streamlines of the electric polarization **P** form closed loops centered along a straight line (coinciding with the polarization of the incident magnetic field) that generate a finite magnetic dipole moment[11]. Physically, the difference between these Mie-like modes can be explained in terms of ordinary and extraordinary electric polarizations, which are distinguished by the fact that **E** has *z*- and *x*-(or *y*-) components or purely lies in (*x*,*y*)-plane, respectfully (Fig. 1). For calcite the magnetic mode at $\lambda = 556$ nm is purely ordinary owing to vanishing $E_z$, while finite components of both $E_x$ and $E_z$ make the magnetic mode at $\lambda = 478$ nm to be purely extraordinary. In contrast, for vaterite the ordinary magnetic mode appears at the shorter wavelength ($\lambda = 480$ nm) than extraordinary ($\lambda = 555$ nm). We attribute this discrepancy to the fact that calcite is a negative uniaxial crystal whereas the cylindrical vaterite core, which is responsible for the spherulite Cartesian anisotropy, behaves as an effective positive uniaxial crystal (Fig. 2 (a) – (c)).

Additionally, calcite supports a new hybrid magnetic resonance at $\lambda = 411$ nm, which possesses closed loops for streamlines of **P** centered along an S-shaped curve, which does not have any analogies in the Mie-type solutions (Fig. 4 (a)). As a result, this mode has all three nontrivial components of **E** and cannot be factorized by polarization. In the bulk uniaxial crystal such a field distribution corresponds to an extraordinary wave. However, peculiar coupling between ordinary and extraordinary polarized fields due to their interaction on the curved particle surface yields this hybrid S-shaped mode, which cannot be treated as purely ordinary or extraordinary. Also, its scattering signatures, analyzed in Fig. 3 (b), show that this mode contains significant contributions from spherical electric and magnetic quadrupoles.

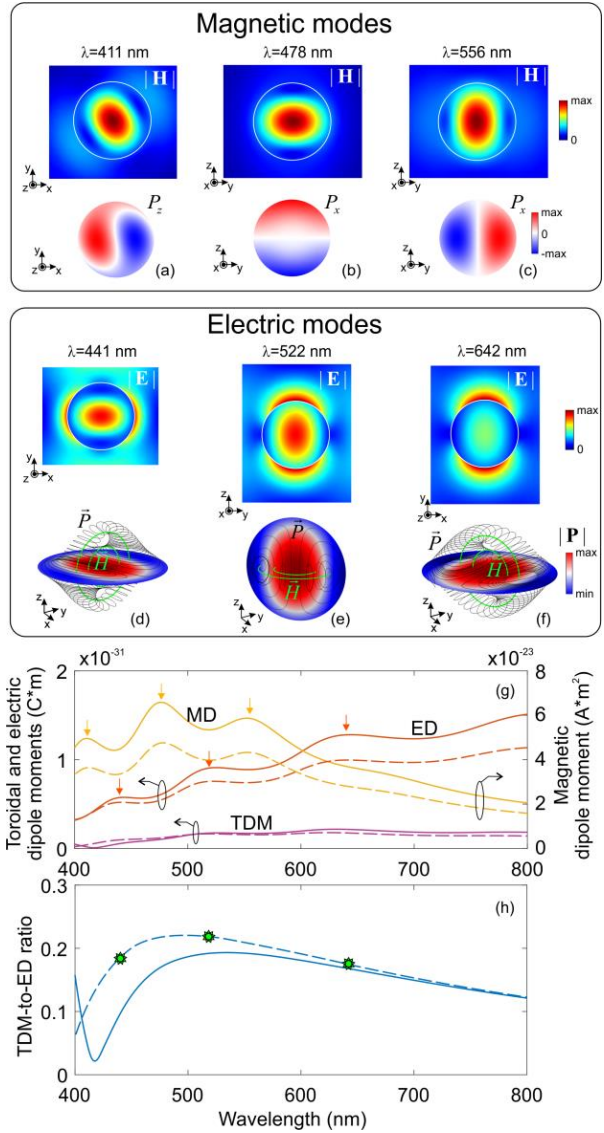

**Fig. 4:** Analysis of magnetic and electric eigen modes for the calcite particle. In (a)-(f) each resonance was obtained with the numerical eigen mode solver (i.e., in the absence of the external field) and characterized in terms of magnetic **H** or electric **E** field and electric polarization **P**. The cross-sections are taken at the particle's center (($y$,$z$) plane). Bottom row of figures in panels (a), (b), and (c) show the color map of corresponding components of **P**; while (d), (e) and (f) show its absolute value along with streamlines for **P** (black) and **H** (green). (g) The spectra of Cartesian multipoles extracted from the scattering problem, whose scattering cross-sections are presented in Fig. 2 (b). Multipoles include electric dipole $|\mathbf{p}_{car}|$ (ED), toroidal dipole $|ik\varepsilon_h \mathbf{T}_{car}/c|$ (TDM),

and magnetic dipole $|\mathbf{m}_{car}|$ (MD). The electric field amplitude of the incident wave is 1 V/m, values of electrical moments – left vertical axis, magnetic moments – right vertical axis. Color bars are provided in linear scale. Arrows indicate peaks corresponding to the eigenmodes shown in (a)-(f). (h) Toroidal-dipole-to-electric-dipole (TDM-to-ED) ratio $|ik\varepsilon_h \mathbf{T}_{car}/c|/|\mathbf{p}_{car}|$. The green stars denote the TDM-to-ED ratios calculated for the field distributions at electric dipole resonant *eigen mode*, shown in (d), (e) and (f). Continuous and dashed curves correspond to parallel and perpendicular orientation of the electric field polarization with respect to *z*-axis.

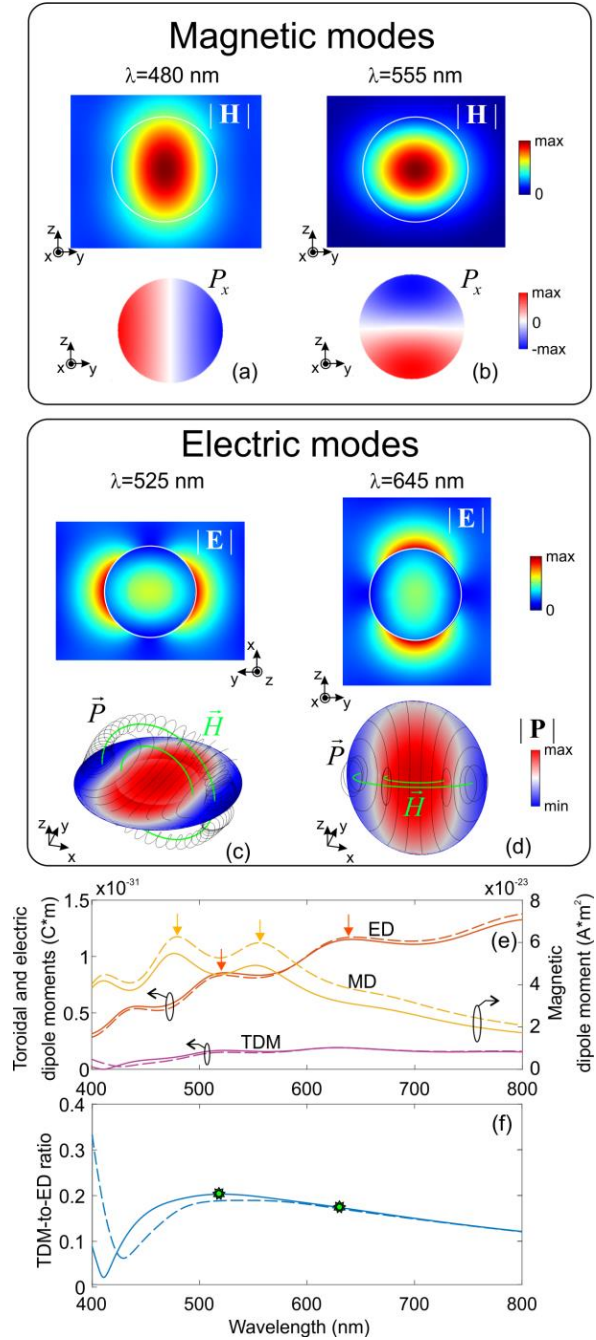

**Fig. 5:** The same caption as in Fig. 4 for the vaterite nanoparticle.

Although spherical multipole decomposition of the scattering shows that electrical modes are associated with the electric dipole (Figs. 3 (b) and (c)), the eigenmode analysis reveals that their structure is significantly different from the Mie-type electric dipoles, which are characterized by the open electric streamlines converging towards two well-defined poles[11]. Specifically, the absolute value of **P** has a single lobe profile for of the all discovered resonances, but its streamlines can form either closed loops or helicoidal-like curves (Figs. 4 (d), (e) and (f) and Figs. 5 (c) and (d)). Such field structures resemble the toroidal dipole, associated with the current of a solenoid, which is bent into a torus[20,21,58,59]. This type of moment cannot be described by any single spherical multipole, but it can be factorized in the total Cartesian representation as [60]

$$\mathbf{P}_{tot,car} = \mathbf{p}_{car} + i\frac{k\varepsilon_h}{c}\mathbf{T}_{car}, \tag{4}$$

where

$$\mathbf{p}_{car} = \iiint \mathbf{P}(r)dV$$

is the ordinary electric dipole moment (ED) and

$$\mathbf{T}_{car} = \frac{1}{10}\iiint \left[ (\mathbf{r}\cdot\mathbf{J}(r))\mathbf{r} - 2r^2\mathbf{J}(r) \right]dV$$

is the electric toroidal dipole moment (TDM). Thus, eq. (4) allows calculating the contribution of TDM to $\mathbf{P}_{tot,car}$. The Cartesian magnetic dipole moment (MD) is given by:

$$\mathbf{m}_{car} = \frac{1}{2}\iiint \mathbf{r}\times\mathbf{J}(r)dV.$$

By using these definitions, we extract ED, TDM and MD from the numerical solution of light scattering by calcite and vaterite nanoparticles (Fig. 4 (g) and Fig. 5 (e)). The Cartesian moments also demonstrate resonant behavior with the peaks corresponding to the wavelengths of eigenmodes shown in Fig. 4 and Fig. 5. Remarkably, ED and TDM are much less sensitive to the polarization of the incident light excitation than MD for both calcite and vaterite. It can be attributed to differences in overlapping between the plane wave and electric and magnetic resonances fields, since ordinary and extraordinary magnetic modes have only 2 nontrivial components of **E** field, while electric resonances are characterized by all 3 non-vanishing components of **E**.

These calculations also confirm the significant contribution of TDM to $\mathbf{P}_{tot,car}$ over the full visible spectral range, reaching 25 % (Figs. 4 (h) and 5 (f)). Remarkably, the spectral peaks of TDM and ED exactly coincide. Additionally, the extraction of ED and TDM contributions from the field distribution of the electric eigenmodes gives the ratio of TDM and ED contributions to $\mathbf{P}_{tot,car}$ (the TDM-to-ED ratio) $|ik\varepsilon_h\mathbf{T}_{car}/c|/|\mathbf{p}_{car}|$ similar to those found for the scattering problem and indicated by green stars in Figs. 4 (h) and 5 (f). This fact underlines the

complex nature of new non-Mie eigenmodes of the particles, which are the nontrivial mixture of different multipoles.

It is worth noting that both vaterite and calcite nanoparticles demonstrate similar anisotropy-driven resonances of both magnetic and electric type. However, for the calcite nanoparticle they are more pronounced in scattering spectrum and are better localized (compare Figs. 3 (b) and (c) and Figs. 4 and 5). This fact results from the smaller effective volume of the vaterite nanoparticle, where Cartesian anisotropy is obtained (the rest of the particle's volume is occupied by a radial-type anisotropy) (Fig. 2). Since the discovered resonances are definitely induced by the Cartesian particle anisotropy, all previous studies considering particles with radial anisotropy (i.e., with different radial and azimuthal/polar components of the permittivity tensor) have never demonstrated any novel families of eigenmodes, except for those given by a slightly modified Mie solution[47,48].

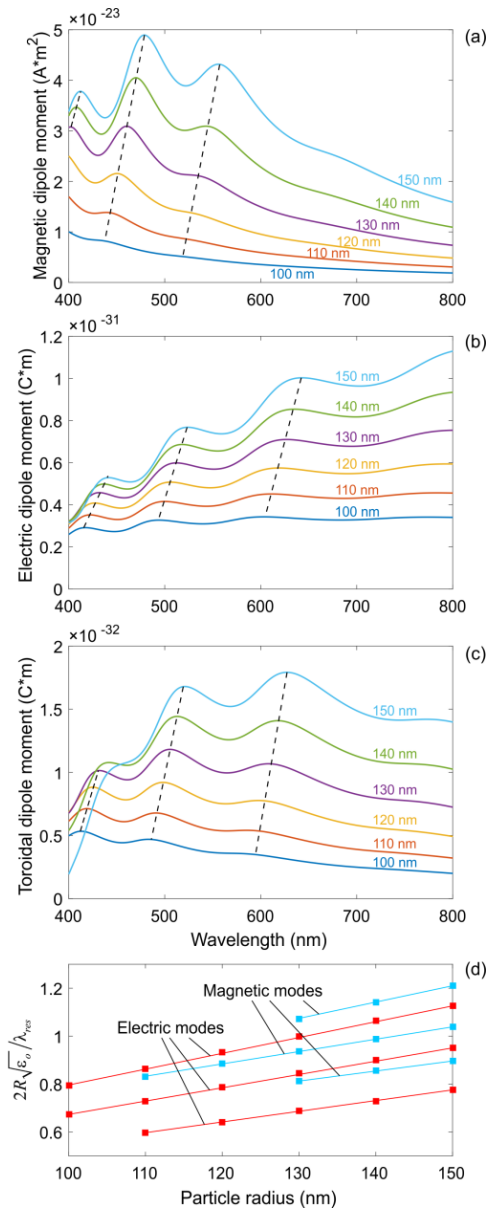

**Fig. 6:** Scaling of non-Mie resonances with variations in the particle size. The spectra of (a) MD $|\mathbf{m}_{car}|$, (b) ED $|\mathbf{p}_{car}|$, and (c) TDM $|ik\varepsilon_h\mathbf{T}_{car}/c|$ for the calcite nanoparticles with different sizes at $\alpha = \pi/2$. A number above each line indicates the particle radius. The electric field amplitude of the incident wave is 1 V/m. Dashed lines highlight linear trends in spectral peaks shift due to the growth of the particle size. (d) The ratio of the particle size to the wavelength of the ordinary wave inside the particle $2R/(\lambda_{res}/\sqrt{\varepsilon_o})$ as a function of the particle radius for the magnetic and electric resonances in (a) – (c). Squares show the data from (a) – (c), and lines mark the linear fits. Resonance peaks of ED and TDM coincide.

Finally, we consider the scaling of non-Mie resonances with variations in the particle size. To this end, we simulate light scattering by calcite spheres at $\alpha = \pi/2$. As follows from Figs. 4(g) and 5(g), all conclusions for this case can be extended to the counterpart field polarization as well as the vaterite nanoparticles. Figures 6 (a) – (c) show gradual growth of the resonances in the spectra of MD, ED and TDM as the particle radius increases from 100 nm to 150 nm. We notice the pronounced linear trends in the spectral shifts of the resonant peaks that is valid for conventional Mie resonances as well. However, there are also several important discrepancies. The MD Mie mode always has the lowest frequency among other Mie modes, and for this resonance the ratio of the particle size to the wavelength inside the structure is close to unity[10]. For the ED Mie mode this size parameter is close to 1.34. In contrast, for the non-Mie eigenmode family the ED mode is of the lowest frequency. Moreover, both ED and MD non-Mie resonances do exist when the size parameter is significantly below 1, as depicted in Fig. 6 (d). This is an additional clear evidence that non-Mie resonances represent a novel family of optical modes where optical field confinement is reached through anisotropy rather than due to a high refractive index.

The observation of non-Mie resonances in experiment can be performed by standard dark-field microscopy[29,31,58]. The protocols for preparation of spherical vaterite and calcite nanoparticles were reported in Refs. [51,61]. The orientation of an anisotropic nanoparticle can be fixed in any desirable direction by using an optical tweezer in addition to the probing light beam as the particle tends to align its optical axis with the electric field of the tweezer beam, as demonstrated in Refs.[42,43]. The spectral mode behavior in experiment will depend on the refractive indexes of a substrate and a surrounding media. In case of the glass substrate and the air environment, we would not expect significant broadening and shifting of the spectral peaks because of quite a large contrast with the particle refractive indexes. However, the drop in the contrast (for example, for the substrate of indium tin oxide) would result in worse optical field bounding and peaks broadening.

## CONCLUSION AND OUTLOOK

In summary, we revealed a novel family of optical resonances that exist in low-index spherical nanoparticles due to Cartesian anisotropy. The numerical simulations of light scattering by calcite and vaterite nanoparticles along with the eigenmode analysis allowed to identify the magnetic dipole resonances of ordinary, extraordinary and hybrid polarizations as well as intricate electric dipole resonances which feature a significant toroidal electric dipole moment. These results clearly show that optical magnetism, subwavelength confinement of light field, and the rich physics of toroidal moments can be employed in living organisms by using biominerals[35,36]. Additionally, calcite and vaterite offer heating-free non-Mie resonances tunable in the full optical domain that yields promising opportunities for quenching-free enhancement of luminescence from the attached emitting systems (e.g., dyes, quantum dots, etc.)[62].

Our results also provide interesting novel insights into interactions of living organisms with light. Brain photobiomodulation therapy, an innovative treatment for a wide range of neurological and psychological conditions, employs optical waves to stimulate desired neural activity[63]. Particularly, it has been shown that seasonal affective disorder can be cured by delivering light into brain through ear canals[64]. However, the physical mechanism behind this process has remained unrevealed owing to complicated structure of curved ear canals, resulting in challenges in direct stimulation of neurons. Taking in account our findings, one may hypothesize that otoliths (biomineral particles made of calcite), residing in the vestibular system in the end of ear canals and reaching a micron size, can support non-Mie optical resonances, giving rise to efficient stimulation of nerves by light.

## ACKNOWLEDGEMENTS

The research was supported in part by ERC StG 'In Motion', PAZY Foundation (Grant No. 01021248), and Tel Aviv University Breakthrough Innovative Research Grant. REN acknowledges A. S. Shalin and A. Machnev for fruitful discussions.